          [1999/12/02 1.01 (PWD)]
\documentclass[a4paper,twocolumn]{esapub}
\usepackage{graphics}

\usepackage{natbib}
\newcommand{\be}{\begin{equation}}
\newcommand{\ee}{\end{equation}}
\title{Variable Unidentified Gamma-Ray Sources Near the Galactic Plane}

\author{Gustavo E. Romero}
\author{Diego F. Torres}
\author{P. Benaglia}
\author{J.A. Combi}
\affil{Instituto Argentino de Radioastronom\'{\i}a, C.C. 5, 1894
Villa Elisa, Argentina}
\author{B. Punsly}
\affil{4014 Emerald Street, No. 116, Torrance, CA 90503, USA}

\begin{document}


\maketitle

\begin{abstract}
In a recent paper, Romero et al. (1999) have presented the results
of a positional correlation analysis of the low-latitude
unidentified gamma-ray sources in the Third EGRET catalog with
complete lists of galactic objects like supernova remnants, early
type stars with strong stellar winds, and OB associations
(considered as pulsar tracers). We have now carried out a study of
those 42 sources at $|b|<10^{\circ}$ for which no counterpart was
found at all. A variability analysis shows that this sample
contains a population of sources with high levels of variability
at gamma-rays. The surface density of these variable sources is 5
times higher than what is expected for unidentified AGNs seen
through the Galactic plane. We discuss the origin of this presumed
Galactic population of gamma-ray objects and the role that
INTEGRAL could play in their physical identification.

\end{abstract}

\section{Introduction}

The Third EGRET (3EG) catalog of high-energy gamma-ray sources
lists 271 point-like sources detected at $E>100$ MeV (Hartman et
al. 1999). Among these sources there are 66 high-confidence
identifications with blazars, 7 confirmed gamma-ray pulsars, and
about 170 unidentified sources. The distribution of the
unidentified sources with Galactic latitude shows a strong density
gradient towards the Galactic plane, which indicates a strong
contribution from sources belonging to our Galaxy (see Figure 1).
Additionally, a group of mid-latitude sources has been recently
found to be correlated with the Gould belt, a nearby region of
active star formation (Grenier \& Perrot 1999, Gehrels et al.
2000).

The positional correlation between the 3EG sources in the Galactic
plane and different types of Galactic objects was studied by
Romero et al. (1999). They found that 10 sources are coincident,
within the positional uncertainties, with Wolf-Rayet and Of stars
(which are hot and massive stars endowed with strong supersonic
stellar winds), 26 with OB associations (usually considered as
pulsar tracers) and 22 with supernova remnants (SNRs). The
probability of pure chance superposition with OB associations and
SNRs is quite negligible ($<10^{-5}$). For the stars, the case is
suggestive although not so conclusive (probabilities $<10^{-2}$).
Romero et al. (1999) also found a set of 42 gamma-ray sources for
which there is not positional superposition with any known
Galactic object capable to produce high-energy gamma-rays. The
nature of this sub-group of sources remains a mystery.

In this paper we present a variability analysis of the sample of
unidentified 3EG sources near the Galactic plane that do not
present positional correlation with potential Galactic
counterparts. Such an analysis can shed some light on the nature
of the parent population, because radio-quiet pulsars and yet
undiscovered SNRs are not expected to present high levels of
variability on timescales of months. Based upon the results, we
shall advance a working hypothesis on the parent objects.

\section{Variability analysis}

There are 81 unidentified sources at $|b|<10^{\circ}$ listed in
the 3EG catalog. We shall focus our analysis on the sample of 40
sources with no known counterpart.\footnote{We have excluded 2
sources that are almost certainly artifacts associated with the
proximity of the very bright Vela pulsar (these sources do not
show up in a map which excludes the Vela pulsation intervals) and
hence the difference with the number of 42 sources mentioned by
Romero et al. (1999). There are other 4 possible artifacts that
are positionally correlated with SNRs.}

\begin{figure}[t]\vspace{-3cm}
\begin{flushleft}
\resizebox{7cm}{!} {\includegraphics{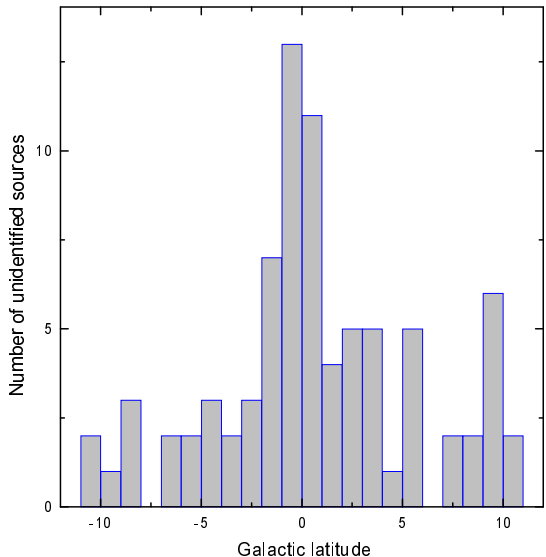}}
\vspace{-1.5cm} \caption{Distribution of unidentified gamma-ray
sources with Galactic latitude, for the interval
$|b|<10^{\circ}$.\label{fig1}}
\end{flushleft}
\end{figure}

Zhang et al. (2000) have already presented a variability analysis
of the sources that are correlated with OB associations and SNRs,
finding that most of these sources could be non-variable. These
authors have shown that this group of sources could be mostly
originated in a population of young and/or radio-quiet gamma-ray
pulsars. We aim here at establishing the variability properties of
those sources not included in Zhang et al. (2000) analysis.

In order to carry out the variability analysis we define a mean
weighted value for the EGRET flux as: \be \left< F \right> =
\left[ \sum_{i=1}^{N_{{\rm vp}}} \frac{F(i)}{\epsilon(i)^2}
\right]\times \left[ \sum_{i=1}^{N_{{\rm vp}}} \frac
1{\epsilon(i)^{2}} \right]^{-1}.\ee Here, $N_{{\rm vp}}$ is the
number of viewing periods for each gamma-ray source. We take into
account only single viewing periods. $F(i)$ is the observed flux
in the $i$-period, whereas $\epsilon(i)$ is the corresponding
error. For those observations in which the significance
($\sqrt{TS}$ in the EGRET catalog) is greater than 3$\sigma$, we
took the error as $\epsilon(i) = F(i)/\sqrt{TS}$. However, many of
the observations are in fact upper bounds on the flux, with
significance below 2$\sigma$. For these ones, we assume both
$F(i)$ and $\epsilon(i)$ as half the value of the upper bound.
This is a rather conservative assumption that is not expected to
result in an overestimate of the variability levels, as it could be
the case if zero flux is directly adopted. We
then define the fluctuation index $\mu$ as: \be \mu =100\times
\sigma_{{\rm sd}}\times \left< F \right>^{-1} .\ee In this
expression, $\sigma_{{\rm sd}}$ is the standard deviation of the
flux measurements, taking into account the previous
considerations.

In order to remove as far as possible spurious variability
introduced by the observing system, we computed the fluctuation
index $\mu$ for the 7 confirmed gamma-ray pulsars (i.e. those
listed in the 3EG catalog plus the sources 3EG 0634+0521 and 3EG
1048-5840, which were recently identified by Cusumano et al. 2000
and Kaspi et al. 2000, respectively). We adopt the physical
criterion that pulsars are --i.e. by definition-- non-variable
gamma-ray sources. Then, any non-null $\mu$-value for pulsars is
attributed to experimental uncertainty. We then define a
statistical index of variability, $I$, as \be I=\frac{\mu_{{\rm
source}}}{<\mu>_{{\rm pulsars}} }. \ee Then, non-variable sources
are defined as those for which $I<1+1\sigma$. Sources with $I>1$
at a $3\sigma$ level are classified as variable ($1\sigma=0.5$).
Sources with $I>1$ at less than $3\sigma$ are dubious cases and we
cannot conclude about their variability within the present
observational accuracy.

We adopt this criterion, previously used in blazar variability
analysis (e.g. Romero et al. 1994), instead of the similar one
used by McLaughlin et al. (1996) in order to have a direct
comparison with the spurious statistical variability shown by
pulsars. Since the $I$-index establishes how variable is a source
with respect to the pulsar population, it can be considered as
indicative of how confident we can be about the possible physical
variation in the gamma emission of the sources.

\section{Results}

The result of our analysis is presented in Figure 2 in the form of
an histogram. We have fitted the distribution with a Gaussian
curve in order to estimate the position of the maximum. The peak
of the histogram occurs at $I=2.0$, which means that there is an
important contribution from sources that display significant
variability over timescales of months. About 30 \% of the sample
(12 sources) presents indices above 2.5, which are clearly
variable at more than a 3$\sigma$ confidence level. There are
sources with indices as high as $I=8.9$, like 3EG J1735-1500.
There are also 18 dubious sources (45 \%) with $1.5<I<2.5$. Only
10 out of 40 sources (25 \%) are clearly non-variable. The
variability histogram for the sources in the sample is very
similar to that obtained for AGNs in the 3EG catalog (see Figure
3), which is a well known variable population.

Our results are, in general terms, similar to those recently presented by Tompkins (1999),
who re-analyzed the EGRET data to take into account not only all sources included
in the 3EG catalog, but also the 145 marginal detections not included in the final
official list. Two of the sources pointed out by Tompkins as the most variable ones at low
latitudes are included in our sample with indices of 3 and 5.3. Although the dubious cases trend to be non
variable in the most comprehensive Tompkins´ analysis, our method is, we think, still the best
approach when data from the published catalog are used.

\begin{figure}[t] \vspace{-2.9cm}
\begin{flushleft} \resizebox{8.0cm}{!}{\includegraphics{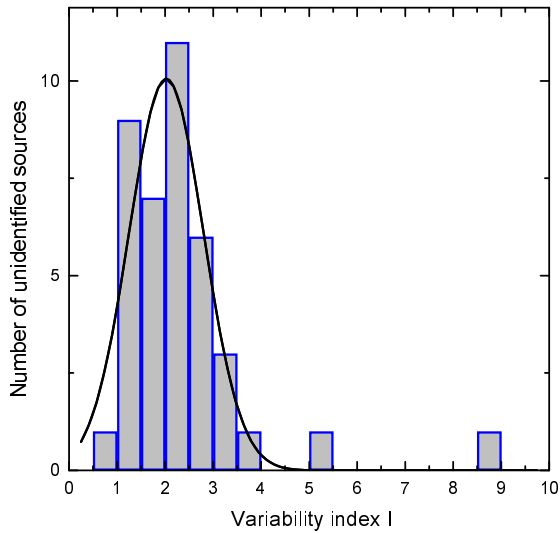}}
\vspace{-1.5cm}\caption{Distribution with variability index for
sources that are not correlated with known Galactic
objects.\label{fig3}} \end{flushleft}
\end{figure}

\begin{figure} [t]\vspace{-2.5cm}
\centering \resizebox{8.0cm}{!}{\includegraphics{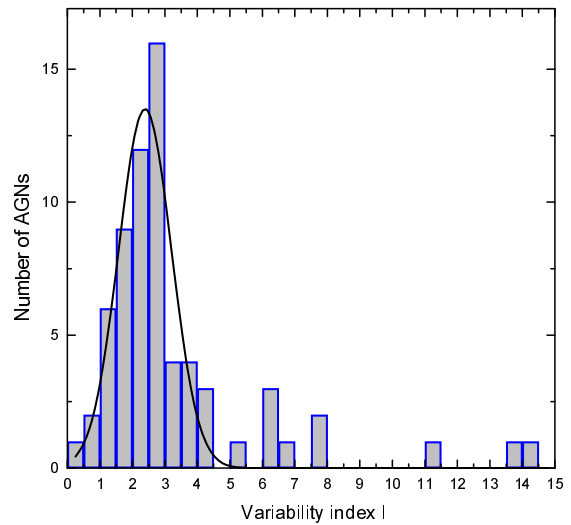}}
\vspace{-1.9cm} \caption{Distribution with variability index for
sources that are identified with AGNs in the 3EG
catalog.\label{fig4}}
\end{figure}

In Figure 4 we present a plot of the variability versus the photon
spectral index $\Gamma$ (defined such that $F(E)\propto
E^{-\Gamma}$) for sources in our sample. In order to provide a
comparison we have also included in this plot the population of
known gamma-ray pulsars and the population of identified AGNs. All
known gamma-ray pulsars are located between the solid horizontal
lines. It is interesting to note that those unidentified sources
with the steepest spectra seem to have the highest variability indices.
The probability of this effect being the effect of chance is
$\sim 8$ \%, not too low, but sufficient as to encourage further studies.

\begin{figure} [t]\vspace{-.5cm}
\begin{flushleft}
\resizebox{8.5cm}{!} {\includegraphics{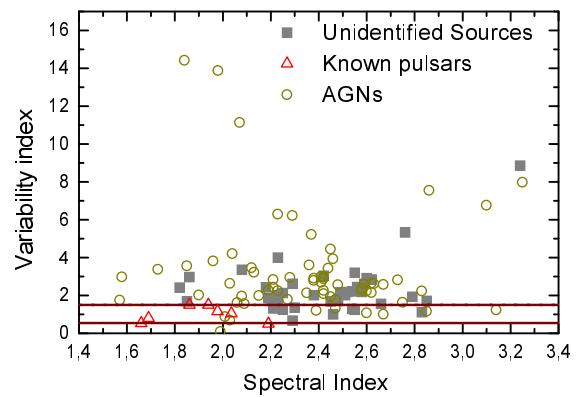}}
\vspace{0.cm} \vspace{-6.5cm} \caption{Variability vs. photon
spectral index for the sources under study. Known gamma-ray
pulsars and identified AGNs are also shown for
comparison.\label{fig5}}
\end{flushleft}
\end{figure}

\begin{figure} [t]\vspace{-.5cm}
\begin{flushleft}
\resizebox{6.5cm}{!} {\includegraphics{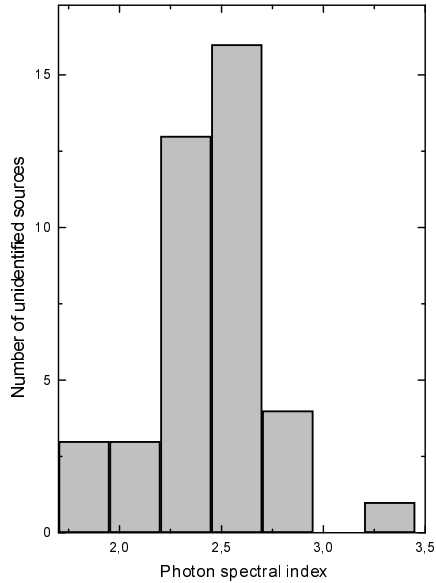}}
\vspace{0.cm} \vspace{-0.5cm} \caption{Histogram with the
distribution of the high-energy photon index for the sources
included in the present study. Notice that most sources have
steeper indices than those presented by known gamma-ray pulsars.
\label{fig6}}
\end{flushleft}
\end{figure}

Figure 5 shows the histogram with distribution of the high-energy
photon index $\Gamma$ for the sources of our sample. Notice that
known gamma-ray pulsars have hard spectra at $E>100$ MeV, with
indices typically $\Gamma<2.15$ (e.g. Thompson et al. 1994),
contrary to most of the sources here under discussion, which
present an average index $<\Gamma>=2.4\pm0.2$.

The nature of the variable population of unidentified sources is a
very intriguing problem that might be addressed by the INTEGRAL
mission.

\section{Magnetized black holes}

The variability analysis presented in the previous section clearly
shows the existence of a galactic populations of variable
gamma-ray sources that do not show up significant emission at
other wavelenghts. The surface density of variable sources in the
sample here under study is five times higher than what is expected
for unidentified AGNs seen through the Galactic plane.

Recently, Punsly et al. (2000) proposed that variable galactic
sources, which from its gamma-ray properties resemble very much
AGNs, could be isolated Kerr-Newman black holes. Punsly (1998) has
shown that a magnetospheric charge may be supported in a disk or
ring around a charged, rotating black hole of a few solar masses.
The total configuration is stable and has zero net charge, hence
the hole should not discharge quickly through accretion from the
diffuse ISM. Unlike the case of pulsars, magnetized black holes do
not present solid surfaces so they do not produce thermal X-ray
emission. Strong magnetized bipolar winds are expected in the form
of collimated electron-positron jets (Punsly 1998), where
gamma-ray emission is generated through inverse Compton scattering
of synchrotron photons and pair annihilations.

Detailed calculations of the spectral energy distribution of these
objects (Punsly et al. 2000, Combi et al. in these proceedings)
show that at MeV energies the annihilation luminosity exceeds the
self-Compton luminosity, producing a break in the spectrum and a
steepening at high-energies. The sources with the steepest
gamma-ray spectra are those whose jets have the highest pair
number densities. Dense jets, in turn, are prone to undergo plasma
instabilities. Synchrotron cooling will most likely cause
radiative losses of the random transverse (thermal) energy in a
magnetized jet. One of the instabilities resulting from the
expected pressure anisotropy in a jet is associated with the shear
Alfv\'en wave: the firehose instability. The fastest growing modes
of this instability have a growth rate $\propto n^{1/2}$, where
$n$ is the pair density. Hence, very dense jets with large
annihilation luminosities and a steep spectral indices tend to be
very firehose unstable. The wobble implies variations of the
Doppler factor of the beamed emission, $\delta$, and consequently
of the luminosity of the jet, which has a dependency
$L\propto\delta^{3}$. Hence, those sources with the steepest
gamma-ray indices would display the highest variability indices,
as observed in our sample of unidentified gamma-ray sources with
no positional counterpart at lower frequencies.

\section{Additional comments}

The magnetized black hole hypothesis for the parent population of
the low-latitude, variable gamma-ray sources can be tested by the
IBIS instrument of INTEGRAL mission. The spectra of these objects
should exhibit a broad peak at a few MeV, similar to that observed
in MeV blazars (e.g. Bloemen et al. 1995). The expected luminosity
of this peak is $\sim 10^{35}$ erg s$^{-1}$ (Combi et al., these
proceedings), so if it occurs at, say, 4 MeV in a source at $\sim
1$ kpc, it should be clearly detected (see Combi et al., these
proceedings). The exact position of the peak in the spectral
energy distribution can be used to infer the Doppler factor of the
jets, whereas its measured intensity can be applied to the
calculation of the pair density in the relativistic flow. The
annihilation luminosity is given by (Punsly et al. 2000):
\begin{eqnarray}
 L_0^{\rm ann}&=&(3/32) \sigma_T
c (\Gamma-1)^2 n^2 m_e c^2 V\times \nonumber\\ &&
\left[\frac{2}{(\Gamma-1/2)(\Gamma+1/2)}+\frac{2(2\Gamma-1)}{\Gamma^2
(\Gamma-1)^2 }\right], \label{ann}
\end{eqnarray}
where $\sigma_T$ is the Thomson cross section, $V$ is the volume
where the annihilations occur, $n$ is as before the number density
of electron-positron pairs, $\Gamma$ is the high-energy photon
index, and $m_e$ is the electron rest mass.

Frequent observations of these sources with IBIS will be useful to
establish the short-term variability, which in turn can shed light
on the involved plasma instabilities. High-resolution radio
observations towards those sources detected by IBIS could find the
synchrotron radio signature of the jets, that should appear as a
point-like non-thermal source of a few tens of mJy at 5 GHz
(Punsly et al. 2000).

\section{Conclusions}

In this paper we have estimated the variability of a sample of 40
low-latitude, unidentified gamma-ray sources that are not
positionally correlated with known potential Galactic emitters. We
have shown that within this set there exists a group of sources
which display high levels of variability over timescales of months
to years at gamma-rays. These sources are unlikely to be pulsars
or undiscovered SNRs. The INTEGRAL satellite will provide a
powerful tool through its IBIS imager to probe the nature of this
population of energetic objects and test whether the hypothesis
here presented, namely that the gamma-ray emission is produced by
isolated magnetized black holes, is correct or not.

\subsection*{Acknowledgments}

This work has been supported by CONICET, ANPCT (PICT No. 03-04881)
and Fundaci\'on Antorchas. G.E.R. is very grateful to the
organizers for a travel grant that made possible his participation
in the Workshop.


\end{document}